 \documentclass[final,5p,times,twocolumn]{elsarticle}

 \usepackage{graphicx}
\usepackage{amssymb}
\usepackage{amsmath}
\journal{Journal of Magnetism and Magnetic Materials}

\begin{document}

\begin{frontmatter}

% Title, authors and addresses

% use the tnoteref command within \title for footnotes;
% use the tnotetext command for theassociated footnote;
% use the fnref command within \author or \address for footnotes;
% use the fntext command for theassociated footnote;
% use the corref command within \author for corresponding author footnotes;
% use the cortext command for theassociated footnote;
% use the ead command for the email address,
% and the form \ead[url] for the home page:
% \title{Title\tnoteref{label1}}
% \tnotetext[label1]{}
% \author{Name\corref{cor1}\fnref{label2}}
% \ead{email address}
% \ead[url]{home page}
% \fntext[label2]{}
% \cortext[cor1]{}
% \address{Address\fnref{label3}}
% \fntext[label3]{}

\title{Scaling form of zero-field-cooled and field-cooled susceptibility in superparamagnet}

% use optional labels to link authors explicitly to addresses:
% \author[label1,label2]{}
% \address[label1]{}
% \address[label2]{}

\author{Masatsugu Suzuki, Sharbani I. Fullem, and Itsuko S. Suzuki}

\address{Department of Physics, State University of New York at Binghamton, Binghamton, New York 13902-6000}

\begin{abstract}
% Text of abstract
The scaling form of the normalized ZFC and FC susceptibility of superparamagnets (SPM's) is presented as a function of the normalized temperature $y$ ($=k_{B}T/K_{u}\langle V\rangle$), normalized magnetic field $h$ ($=H/H_{K}$), and the width $\sigma$ of the log-normal distribution of the volumes of nanoparticles, based on the superparamagnetic blocking model with no interaction between the nanoparticles. Here $\langle V\rangle$ is the average volume, $K_{u}$ is the anisotropy energy, and $H_{K}$ is the anisotropy field. Main features of the experimental results reported in many SPM's can be well explained in terms of the present model. The normalized FC susceptibility increases monotonically increases as the normalized temperature $y$ decreases. The normalized ZFC susceptibility exhibits a peak at the normalized blocking temperature $y_{b}$ ($=k_{B}T_{b}/K_{u}\langle V\rangle$), forming the $y_{b}$ vs $h$ diagram. For large $\sigma$ ($\sigma >0.4$), $y_{b}$ starts to increase with increasing $h$, showing a peak at $h=h_{b}$, and decreases with further increasing $h$. The maximum of $y_{b}$ at $h=h_{b}$ is due to the nonlinearity of the Langevin function. For small $\sigma$, $y_{b}$ monotonically decreases with increasing $h$. The derivative of the normalized FC magnetization with respect to $h$ shows a peak at $h$ = 0 for small $y$. This is closely related to the pinched form of $M_{FC}$ vs $H$ curve around $H$ = 0 observed in SPM's.
\end{abstract}

\begin{keyword}
% keywords here, in the form: keyword \sep keyword
superparamagnet \sep blocked state
% PACS codes here, in the form: \PACS code \sep code
\PACS 75.10.Nr \sep 75.30.Cr \sep 75.50.Tt
% MSC codes here, in the form: \MSC code \sep code
% or \MSC[2008] code \sep code (2000 is the default)

\end{keyword}

\end{frontmatter}

% main text
\section{\label{intro}Introduction}
It is well known that the magnetic susceptibility of superparamagnets (SPM's) and spin glasses (SG's) strongly depends on the way how the cooling of the system is done before or during the measurement \cite{ref01}. Typically there are two types of susceptibility, zero-field-cooled (ZFC) susceptibility and field-cooled (FC) susceptibility. In the case of the ZFC susceptibility ($\chi_{ZFC}$), the system is cooled in the absence of an external magnetic field ($H$) from high temperatures well above a freezing temperature $T_{f}$ (a blocking temperature $T_{b}$ for the SPM's and a spin freezing temperature $T_{g}$ for the SG's) to the lowest temperature well below $T_{f}$ (the ZFC cooling protocol). After $H$ is applied at the lowest temperature, the ZFC susceptibility is measured with increasing temperature ($T$). In contrast, the FC susceptibility ($\chi_{FC}$) is measured in the presence of $H$ with decreasing $T$ from high temperatures well above $T_{f}$ to the lowest temperature well below $T_{f}$ (the FC cooling protocol).

It has been experimentally confirmed that the $T$ dependence of the susceptibility for the SPM's is essentially different from that for the SG's \cite{ref02,ref03,ref04,ref05,ref06,ref07,ref08,ref09,ref10,ref11,ref12}. For SPM's \cite{ref03}, the ZFC susceptibility exhibits a peak at the blocking temperature $T_{b}$, while the FC susceptibility monotonically increases with decreasing $T$, showing no anomaly at $T_{b}$. The difference between the FC and ZFC susceptibility appears well above $T_{b}$. For SG's \cite{ref03}, the ZFC susceptibility also exhibits a peak at the spin freezing temperature $T_{g}$, while the FC susceptibility increases with decreasing $T$ above $T_{g}$ and starts to decrease or to saturate below $T_{g}$, indicating the evidence of glass transition. The difference between the FC and ZFC susceptibility appears just at $T_{g}$.

The purpose of the present paper is to explain the peculiar features on the $T$ and $H$ dependence of the ZFC susceptibility, which were reported in many SPM's \cite{ref02,ref03,ref04,ref05,ref06,ref07,ref08,ref09,ref10,ref11,ref12}. The first feature is as follows. The ZFC susceptibility shows a peak at the blocking temperature $T_{b}$. The blocking temperature $T_{b}$ changes with $H$ in two different ways, depending on the nature of SPM's. (i) $T_{b}$ first increases with increasing $H$, reaching a peak and then decreases with further increasing $H$ (unusual type). (ii) $T_{b}$ monotonically decreases with increasing $H$ (conventional type). The feature (i) is observed for SPM's such as Fe$_{3}$O$_{4}$ nanoparticles (Luo et al. \cite{ref02}), natural horse-spleen (Friedman et al. \cite{ref07}), $\gamma$-Fe$_{2}$O$_{3}$ nanoparticles with diameter 7 nm and ferritin (Sappey et al. \cite{ref08}), and a diluted magnetic fluid composed of FePt nanoparticles (Zheng et al. \cite{ref12}). These two different features are observed even in the same SPM, depending on the conditions of samples (such as dilute and concentrated samples). For $\gamma$-Fe$_{2}$O$_{3}$ nanoparticles in dispersed polymer show that the feature (i) is observed for dilute samples and that the feature (ii) is observed for concentrated samples (Kachkachi et al. \cite{ref11}). The second feature is the $H$ dependence of the FC magnetization $M_{FC}$ near $H$ = 0. The derivative of $M_{FC}$ with respect to $H$ has a sharp peak at $H$ = 0 for ferritin (Tejada et al. \cite{ref09}) and for CoFe$_{2}$O$_{4}$ nanoparticles embedded in potassium silicate (Zhang et al. \cite{ref05}). For natural horse-spleen the magnetic hysteresis shows an anomalous pinched hysteresis loop near zero field (Friedman et al. \cite{ref07}). 

Here we present our numerical analysis of FC and ZFC susceptibility, based on the superparamagnetic blocking model with no interaction between the nanoparticles. The theory used here is the same as that used by Bitto et al. \cite{ref03}, and is essentially the same as the N\'{e}el-Brown model \cite{ref13,ref14}. We show that the main features of the experimental results of the ZFC and FC susceptibility for SPM's can be well explained in terms of the present model. The peculiar features of the susceptibility is due to the Langevin function of the magnetization. Our numerical results will be compared with those predicted by Sappey et al. \cite{ref08}. We show that the blocking temperature $T_{b}$ is extremely sensitive to the width $\sigma$ in the log-norm distribution for volumes of nanoparticles. Resonant tunneling is one of the models to explain the shift of $T_{b}$ with $H$ \cite{ref06,ref15}. For example, $T_{b}$ is 65 K for $\gamma$-Fe$_{2}$O$_{3}$ (Sappy et al. \cite{ref08}) and is too high for the quantum tunneling by which the magnetic moment of the particles can flip. So this possibility may be ruled out. 

\section{\label{back}Scaling form of $\chi_{ZFC}$ and $\chi_{FC}$}
\subsection{\label{back1}Magnetization in the SPM state}
We consider a single-domain particle (nanoparticle) with volume $V$ and the anisotropy having uniaxial symmetry. We assume that there is no interaction between these domains. The simplest form of the anisotropy energy is given by $K_{u}V \sin^{2}\phi$, where $\phi$ is the angle between the magnetization direction and the easy axis, and $K_{u}$ is the anisotropy energy per unit volume. The two states [$\phi = 0$ (partallel) and $\phi=\pi$ (antiparallel)] as the ground state, are energetically degenerate and separated by a energy-barrier height $\Delta E_{a}(V) = K_{u}V$. At temperatures well above $T_{b}$, the magnetization of the single-domain fluctuate due to the thermal activation energy ($k_{B}T$) in a paramagnetic way (SPM state). The definition of $T_{b}$ will be given in Sec.~\ref{back3}. The magnetization (the magnetic moment per unit volume) in the SPM state is described by \cite{ref03}
\begin{equation}
M^{spm}=M_{s}\langle\cos\theta\rangle  ,
\label{eqn01}
\end{equation}
with
\begin{eqnarray}
\langle\cos\theta\rangle
&=&\frac{\int e^{-\frac{U}{k_{B} T}} \cos\theta d\Omega}{\int e^{-\frac{U}{k_{B}T}}d\Omega} 
\nonumber \\
&=&\frac{\int\limits_{0}^{\pi}e^{\frac{M_{s}VH\cos\theta }{k_{B}T}}\cos\theta (2\pi\sin\theta d\theta )}
{\int\limits_{0}^{\pi}e^{\frac{M_{s} VH\cos\theta}{k_{B}T}}(2\pi\sin\theta d\theta)}
\nonumber \\
&=&L(\frac{M_{s} VH}{k_{B} T})
\label{eqn02}
\end{eqnarray}
where $\mu$ ($=M_{s}V$) is the magnetic moment of the single domain with the volume $V$, $M_{s}$ is the saturation magnetization (the magnetic moment per unit volume), $\theta$ ($0\le\theta\le\pi$) is the angle between $\boldsymbol\mu$ and $H$ (the $z$ axis), $\Omega$ is the solid angle, $d\Omega=2\pi\sin\theta d\theta$, $k_{B}$ is the Boltzmann constant, $U$ is the Zeeman energy defined by
\begin{equation}
U=-{\boldsymbol\mu}\cdot{\bf H}=-M_{s}VH\cos\theta
\label{eqn03}
\end{equation}
and $L(\zeta)$ is the Langevin function of $\zeta$;
\begin{equation}
L(\zeta)=\coth (\zeta)-\frac{1}{\zeta} .
\label{eqn04}
\end{equation}

\subsection{\label{back2}Magnetization in the blocked state}
At temperatures well below $T_{b}$, this themal activation energy is too small to cause the barrier hopping between the two states ($\phi = 0$ and $\pi$). Then the direction of the magnetization in the single domain is frozen to one of the two states ($\phi = 0$ and $\phi=\pi$). 

\subsubsection{$H \parallel$ easy axis}
Here we evaluate the ZFC susceptibility below $T_{b}$ in the system where the single domains are randomly oriented \cite{ref16,ref17,ref18,ref19}. First, we consider the case when an external magnetic field $H$ is applied along the easy axis. The system is in thermal equilibrium. The total energy $F_{\parallel}$ consists of the anisotropy energy and the Zeeman energy; 
\begin{equation}
F_{\parallel}=V(K_{u}\sin^{2} \phi-HM_{s}\cos\phi)=K_{u}V(\sin^{2}\phi-2h\cos\phi),
\label{eqn05}
\end{equation}
and the derivative is given by
\begin{equation}
\frac{dF_{\parallel}}{d\phi} =2K_{u}V(\cos\phi+h)\sin\phi ,
\label{eqn06}
\end{equation}
where $\phi$ is the angle between the easy axis and the magnetization direction. The normalized field $h$ is defined by $h=H/H_{K}$, where $H_{K}$ is the anisotropy field and is defined by 
\begin{equation}
H_{K}=2K_{u}/M_{s} .
\label{eqn07}
\end{equation}
The free energy $F_{\parallel}$ has a minimum at $\phi = 0$, and $\pi$, and a maximum at $\cos\phi=-h$. The minimum of the free energy $F_{\parallel}$ is given by 
\begin{equation}
F_{\parallel}^{\min}=K_{u}V(-2h) ,
\label{eqn08}
\end{equation}
at $\phi = 0$ and $\pi$, while the maximum of the free energy is given by
\begin{equation}
F_{\parallel}^{\max}=K_{u} V(h^{2} +1) .
\label{eqn09}
\end{equation}
at $\cos\phi = -h$. The energy-height barrier between two states ($\phi = 0$ and $\pi$) is obtained as
\begin{equation}
\Delta F_{\parallel}=F_{\parallel}^{\max }-F_{\parallel}^{\min }=K_{u}V(1+h)^{2} .
\label{eqn10}
\end{equation}
This means that the energy-barrier height increases with increasing $h$. Then the magnetization along the easy axis ($\phi$ = 0 state) is simply given by
\begin{equation}
M_{\parallel}=M_{s}  .
\label{eqn11}
\end{equation}
The appearance of magnetization along the easy axis below $T_{b}$ is indicative of the freezing of the direction of the magneization along the easy axis. This is one of the important features for the blocked state. The susceptibility along the easy direction is equal to zero; 
\begin{equation}
\chi_{\parallel} =0 .
\label{eqn12}
\end{equation}

\subsubsection{$H \perp$ easy axis}
Next we consider the case when the magnetic field is applied to the hard axis (perpendicular to the easy axis). The total free energy $F_{\perp}$ is given by 
\begin{equation}
F_{\perp}=V(K_{u}\sin^{2}\phi -HM_{s}\sin\phi )=K_{u}V(\sin^{2}\phi -2h\sin\phi) ,
\label{eqn13}
\end{equation}
and the derivative is given by
\begin{equation}
\frac{\partial F_{\bot } }{\partial \phi } =2K_{u} V\cos \phi (\sin \phi -h) ,
\label{eqn14}
\end{equation}
where $\phi$ is the angle between the easy axis and the magnetization direction, and $h$ is smaller than 1. The free energy $F_{\perp}$ has a minimum at sin$\phi=h$. Then the magnetization along the hard axis (the field direction) of the single domain is obtained as 
\begin{equation}
M_{\perp}=M_{s}\sin\phi=M_{s}h=\frac{M_{s}}{H_{K}}H ,
\label{eqn15}
\end{equation}
while the susceptibility along the hard axis (the field direction) is 
\begin{equation}
\chi_{\perp} =\frac{M_{\perp}}{H}=\frac{M_{s}}{H_{K}}=\frac{M_{s}^{2}}{2K_{u}}.
\label{eqn16}
\end{equation}
The minimum value of $F_{\perp}$ is given by
\begin{equation}
F_{\perp}^{\min }=-K_{u}Vh^{2} ,
\label{eqn17}
\end{equation}
at $\sin\phi=h$, while the maximum value of $F_{\perp}$ is given by 
\begin{equation}
F_{\perp}^{\max}=K_{u}V(1-2h) ,
\label{eqn18}
\end{equation}
at $\phi=\pi /2$. Then the energy-barrier height between two states denoted by $\sin\theta=h$, is given by the energy difference $\Delta F_{\perp}$ defined by 
\begin{equation}
\Delta F_{\perp}=F_{\perp}^{\max }-F_{\perp}^{\min}=\Delta E_{a}(V,h)=K_{u}V(1-h)^{2} ,
\label{eqn19}
\end{equation}
indicating that the energy-barrier height $\Delta E_{a}(V,h)$ decreases with the increase of $h$ or with the decrease of $V$.

\begin{figure}
\includegraphics[width=6.0cm]{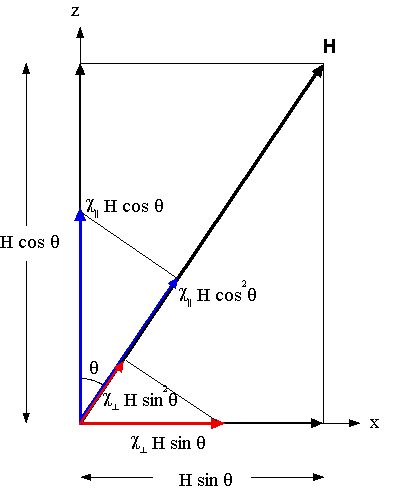}
\caption{\label{fig01}(Color online) The schematic diagram of the easy axis (the $z$ axis) of the single domain and the external magnetic field ${\bf H}$ in the $z$-$x$ plane. The angle between ${\bf H}$ and the $z$ axis is $\theta$. The $z$ and $x$ components of ${\bf H}$ are $H\cos\theta$ and $H\sin\theta$, respectively. Correspondingly, the $z$ and $x$ components of the magnetization induced by the field ${\bf H}$ are  $\chi_{\parallel}H\cos\theta$ and $\chi_{\perp}H\sin\theta$. From the symmetry, only the component of the resultant susceptibility along the direction of ${\bf H}$ is not equal to zero, and is given by $\chi_{\parallel}\cos^{2}\theta+\chi_{\perp}\sin^{2}\theta$.}
\end{figure}

In a system where the direction of the easy axes for single domains is randomly oriented, the average susceptibility (so-called powder susceptibility) can be evaluated as
\begin{eqnarray}
\chi_{av} & = & \chi_{\parallel}\langle\cos^{2}\theta\rangle+\chi_{\perp}\langle\sin^{2}\theta\rangle \nonumber \\
&=&\frac{\int (\chi_{\parallel}\cos^{2}\theta+\chi_{\perp}\sin^{2}\theta) d\Omega}{\int d\Omega}\nonumber \\
& = & \frac{1}{3}\chi_{\parallel}+\frac{2}{3} \chi_{\perp} \nonumber \\
& = & \frac{M_{s}^{2}}{3K_{u}} ,
\label{eqn20}
\end{eqnarray}
using Eqs.(\ref{eqn12}) and (\ref{eqn16}), where $d\Omega = 2\pi\sin\theta d\theta$, $\theta$ ($0\le\theta\le\pi$) is the polar angle between the easy axis (the $z$ axis) and the magnetic-field direction, see Fig.~\ref{fig01} and the caption for the detail of calculation. When the average susceptibility is expressed by Eq.(\ref{eqn20}), we refer to the system as being in the blocked state. This susceptibility is the ZFC susceptibility of the single-domain with volume $V$ below $T_{b}$.

\subsection{\label{back3}Blocking temperature $T_{b}$}
The relaxation time of the magnetization between two states ($\phi$ = 0 and $\pi$ states) is given by thermal activation (Arrehenius law).
In the N\'{e}el-Brown relaxation process \cite{ref13,ref14}, the relaxation time of the magnetization between the two states is given by, 
\begin{equation} 
\tau =\tau _{0} \exp \lbrack \frac{\Delta E_{a}(V)}{k_{B}T}\rbrack,
\label{eqn21} 
\end{equation} 
where $\tau_{0}$ is a microscopic limiting relaxation time (usually $\tau_{0} \approx 10^{-9}$ sec). The measurement time $\tau_{m}$ is typically of the order of $10^{2}$ sec for the DC magnetization measurement. In Eq.~(\ref{eqn21}) with $\tau =\tau_{m} $ and $T=T_{b}(V)$, the blocking temperature $T_{b}(V)$ is derived as 
\begin{equation}
T_{b}(V)=\frac{\Delta E_{a}(V)}{k_{B}\ln ( \tau_{m}/\tau_{0})} .
\label{eqn22}
\end{equation}

In the presence of $H$, it is assumed that the energy-barrier $\Delta E_{a}(V,h)$ is described by
\begin{equation}
\Delta E_{a} (V,h)=K_{u} V( 1-h) ^{\alpha} ,
\label{eqn23}
\end{equation}
where $\alpha = 2$ in the present case. Note that $\alpha = 1.5$ when the orientational order is taken into account \cite{ref08}. The blocking temperature $T_{b}(V,h)$ is given by 
\begin{equation}
T_{b} (V,h)=\frac{K_{u}V}{k_{B}\ln(\tau_{m}/\tau_{0})}(1-h)^{\alpha}.
\label{eqn24}
\end{equation}

\subsection{\label{back4}Form of magnetization in the superparamagnetic state and blocked state}
Before discussing the general case, first we consider a simple case where all nanoparticles have the same volume. Above the blocking temperature $T_{b}(V,h)$, the system is in the SPM state. The magnetization $M^{spm}$ (per unit volume) for the SPM state is given by \cite{ref03} 
\begin{equation}
M_{ZFC}^{spm}(V)=M_{FC}^{spm}(V)=\epsilon M_{s}L(\frac{M_{s}VH}{k_{B} T}) ,
\label{eqn25}
\end{equation}
where $\epsilon$ is the volume fraction occupied by ferromagnetic nanoparticles.

Below $T_{b}(V,h)$, the system is in a blocked state. The magnetization $M^{bl}$ (per unit volume) of the blocked state is given by
\begin{equation}
M_{ZFC}^{bl} (V) =\frac{\epsilon M_{s}^{2} H}{3K_{u} } ,
\label{eqn26}
\end{equation}
and
\begin{equation}
M_{FC}^{bl} (V)=\epsilon M_{s} L( \frac{M_{s} VH}{k_{B} T_{b} (V,h)}
) .
\label{eqn27}
\end{equation}

\subsection{\label{back5}Scaling form of normalized ZFC and FC susceptibility}

\begin{figure}
\includegraphics[width=7.0cm]{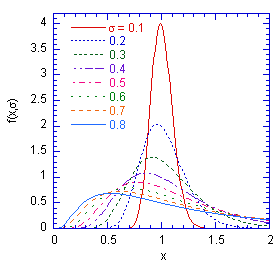}
\caption{\label{fig02}(Color online) Plot of $f(x,\sigma)$ as a function of $x$ at various $\sigma$ ($\sigma = 0.1 - 0.8$). $f(x,\sigma)$ is a log-normal distribution function and is defined by Eq. (\ref{eqn34}).}
\end{figure}

Using Eqs.~(\ref{eqn25})-(\ref{eqn27}), we derive the general expression for the magnetization. We consider a system of ferromagnetic nanoparticles having a wide distribution of volume sizes and let $\langle V \rangle$ be their average volume. We also define a characteristic volume $V_{m}(T,h)$ given by
\begin{equation}
V_{m} (T,h)=\frac{k_{B} T}{K_{u} (1-h)^{\alpha} } \ln (\tau _{m} /\tau _{0} ),
\label{eqn28}
\end{equation}
which is derived from the condition
\begin{equation}
\frac{\Delta E_{a}(V_{m},h)}{k_{B}T}=\ln (\tau_{m} /\tau_{0}) .
\label{eqn29}
\end{equation}
For simplicity we introduce a volume ratio $x$ ($= V/\langle V \rangle$). The characteristic volume ratio $x_{m}$ is then defined by 
\begin{eqnarray}
x_{m} &=&\frac{V_{m} (T,h)}{\langle V\rangle } 
= \frac{k_{B} T}{K_{u} \langle V\rangle (1-h)^{\alpha}}\ln (\tau_{m}/\tau_{0}) \nonumber\\
&=& \frac{y}{(1-h)^{\alpha}}\ln (\tau_{m} /\tau_{0}) ,
\label{eqn30}
\end{eqnarray}
where $y$ is the reduced temperature and is defined by $y = k_{B}T/(K_{u}\langle V\rangle )$. For $x<x_{m}$, the system is in a SPM state, and for $x>x_{m}$, the system is in a blocked state. For the SPM state, the ZFC and FC magnetizations shown in Eq.~(\ref{eqn25}) can be rewritten as 
\begin{equation}
M_{ZFC}^{spm}(x,y,h)=M_{FC}^{spm}(x,y,h)=\epsilon M_{s} L(\frac{2hx}{y}) ,
\label{eqn31}
\end{equation}
while for the blocked state Eqs.~(\ref{eqn26}) and (\ref{eqn27}) are rewritten as
\begin{equation}
M_{ZFC}^{bl} (x,y,h)=\epsilon M_{s} \frac{2h}{3} ,
\label{eqn32}
\end{equation}
and
\begin{equation}
M_{FC}^{bl}(x,y,h)=\epsilon M_{s} L(\frac{2h}{(1-h)^{\alpha} }\ln (\tau_{m}/\tau_{0})) ,
\label{eqn33}
\end{equation}
respectively. 

We assume that volume distribution of the nanoparticles, $f(x,\sigma)$, is expressed by the log-normal function \cite{ref03}
\begin{equation}
f(x,\sigma )=\frac{1}{\sqrt{2\pi } \sigma x} \exp [ -\frac{(\ln
x)^{2} }{2\sigma ^{2} } ] ,
\label{eqn34}
\end{equation}
where
\[
\int\limits_{0}^{\infty }f(x,\sigma )dx =1 , \textrm{ and }
x_{av} =\int\limits_{0}^{\infty }xf(x,\sigma )dx =\exp (\sigma ^{2} /2) .
\]
Here $x_{av}$ is the average of $x$ and $\sigma$ is the width of the distribution. Figure \ref{fig02} shows a plot of $f(x,\sigma)$ as a function of $x$ at various $\sigma$. The log-normal distribution function $f(x,\sigma)$ has a maximum at $x = x_{max}=\exp(-\sigma^{2})$. The value of $x_{max}$ decreases with increasing $\sigma$, while the value of $x_{av}$ increases with increasing $\sigma$. In the limit of $\sigma\rightarrow 0$, $f(x,\sigma)$ becomes a Dirac delta function $\delta (x-1)$ which has a sharp peak at $x$ = 1. Using Eqs.~(\ref{eqn31})--(\ref{eqn34}), the scaling forms of the normalized ZFC and FC susceptibility are given by
\begin{eqnarray}
\frac{\chi _{ZFC} (y,h,\sigma )}{\chi _{0} }
&=&\frac{1}{h} \int\nolimits_{0}^{\infty }[ L( \frac{2hx}{y} ) U_{-1} (x_{m} -x)\nonumber\\
&+&\frac{2h}{3} U_{-1} (x-x_{m} )]  f(x,\sigma )dx ,
\label{eqn35}
\end{eqnarray}
and
\begin{eqnarray}
\frac{\chi _{FC} (y,h,\sigma )}{\chi _{0} }
&=&\frac{1}{h} \int\nolimits_{0}^{\infty }[ L( \frac{2hx}{y} ) U_{-1} (x_{m} -x)\nonumber\\
&+&L( 2h\frac{\ln (\tau _{m} /\tau _{0} )}{(1-h)^{\alpha} } ) U_{-1} (x-x_{m} )]  f(x,\sigma )dx , \nonumber\\
\label{eqn36}
\end{eqnarray}
where $U_{-1}$ is a step function [$U_{-1}(x)=1$ for $x\ge 0$ and 0 for $x<0$]. $\chi_{0}$ [$= \epsilon M_{s}^{2}/(2K_{u})$] is a constant susceptibility and $x_{m}$ given by Eq.~(\ref{eqn30}) is a function of $y$, $h$, and $\tau_{m}/\tau_{0}$. The normalized ZFC and FC susceptibilities depend only on $y$, $h$, $\tau_{m}/\tau_{0}$, and $\sigma$. They are independent of the values of $\epsilon$, $M_{s}$, $K_{u}$, and $\langle V \rangle$. Hereafter, we assume $\ln (\tau_{m}/\tau_{0})=\ln (10^{2}/10^{-9})=25.328$ for simplicity. In this case, $\chi_{ZFC}/\chi_{0}$ and $\chi_{FC}/\chi_{0}$ are described by scaling functions of $y$, $h$, and $\sigma$. 

\section{\label{calc}Numerical calculation}

\begin{figure}
\includegraphics[width=8.0cm]{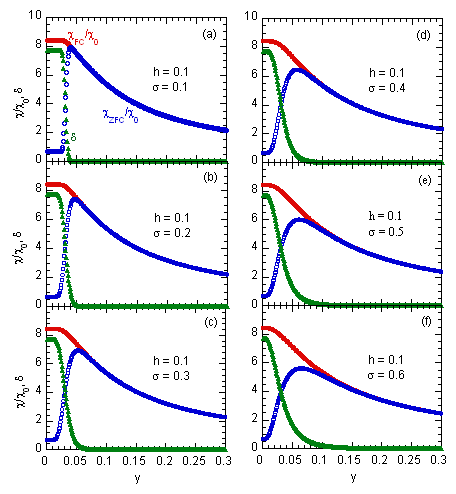}
\caption{\label{fig03}(Color online) Plot of $\chi_{ZFC}/\chi_{0}$, $\chi_{FC}/\chi_{0}$, and $\delta =(\chi_{FC}-\chi _{ZFC})/\chi_{0}$ as a function of reduced temperature $y$. $h$ = 0.1. The width $\sigma$ is varied from 0.1 to 0.6 as a parameter. $\alpha = 2.0$. $\ln (\tau_{m}/\tau_{0})=25.328$. $\chi_{0}=\varepsilon M_{s}^{2}/(2K_{u})$, $y=k_{B}T/(K_{u}\langle V\rangle)$, and $h=H/H_{K}$.}
\end{figure}
 
\begin{figure}
\includegraphics[width=8.0cm]{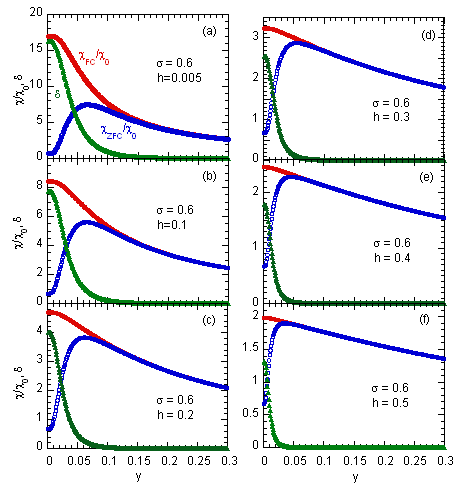}
\caption{\label{fig04}(Color online) Plot of $\chi_{ZFC}/\chi_{0}$, $\chi_{FC}/\chi_{0}$, and $\delta =(\chi_{FC}-\chi_{ZFC})/\chi_{0}$ as a function of reduced temperature $y$. $\sigma$ = 0.6. The reduced field $h$ is changed as a parameter: $h = 0.005 - 0.5$. $\alpha=2.0$}
\end{figure}

Figures \ref{fig03}(a)-(f) show typical plots of $\chi_{ZFC}/\chi_{0}$, $\chi_{FC}/\chi_{0}$ and the difference $\delta$ defined by $\delta =(\chi_{ZFC}-\chi _{FC})/\chi_{0}$ as a function of the normalized temperature $(y=k_{B}T/K_{u} \langle V\rangle )$ , where $\alpha=2.0$ and $\ln (\tau_{m}/\tau_{0}) = 25.328$. The reduced field $h$ ($= H/H_{K})$ is kept constant at $h$ = 0.1, and the width of the log-normal volume distribution function ($\sigma$) is varied as a parameter ($\sigma$ = 0.1 to 0.6). The normalized ZFC susceptibility $\chi_{ZFC} /\chi_{0}$ for $\sigma$ = 0.1 exhibits a peak at the normalized blocking temperature $y_{b}$. The difference $\delta $ starts to appear at the onset normalized temperature of irreversibility $y_{irr}$ (just above $y_{b}$) and increases with decreasing $y$. The irreversible effect of susceptibility occurs below $y_{irr}$. The normalized FC susceptibility $\chi_{FC}/\chi_{0}$ for $\sigma$ = 0.1 is nearly independent of $y$ below $y_{b}$. These features are common to those observed in real SG systems \cite{ref03}. Here, the system behaves like a superspin glass (SSG), where the large magnetic moment of the nanoparticle plays the same role as an atomic spin in a SG system \cite{ref20}. As $\sigma$ increases, the peak of $\chi_{ZFC}/\chi_{0}$ becomes broad and shifts to the large $y$-side. The normalized FC susceptibility $\chi_{FC}/\chi_{0}$ for large $\sigma$ (typically $\sigma$ = 0.6) tends to increase with decreasing $y$ in the small-$y$ region (at low temperatures).

Similar numerical calculations are carried out for the different $\alpha$ ($= 1.5 - 2.0$). We find that the essential points of numerical results are not so different for different $\alpha$. So here we show only the result for $\alpha$ = 2.0. Note that $\alpha$ = 1.5 is used in the paper of Sappey et al. \cite{ref08}.

Figures \ref{fig04} (a)-(f) show typical plots of $\chi_{ZFC}/\chi_{0}$, $\chi_{FC}/\chi_{0}$ and $\delta$ as a function of $y$ for $\sigma=0.6$, where $h$ is changed as a parameter ($h=0.005-0.5$). The normalized susceptibility $\chi_{FC}/\chi_{0}$ at each $h$ increases with decreasing $y$ below $y_{b}$. The normalized susceptibility $\chi_{ZFC}/\chi_{0}$ exhibits a broad peak at $y_{b}$, which shifts to the low-$y$ side with increasing $h$. 

\begin{figure}
\includegraphics[width=7.0cm]{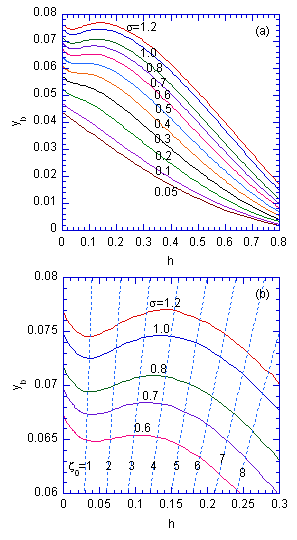}
\caption{\label{fig05}(Color online) (a) $y_{b}$ vs $h$ diagram at various $\sigma$ (= 0.05 - 1.2). $y_{b}$ is the normalized blocking temperature, and $h$ is the normalized field. $\alpha = 2.0$. $\ln (\tau_{m}/\tau_{0})=25.328$. $\sigma$ is the width of the log-norm distribution. (b) Detail of the $y_{b}$ vs $h$ diagram at $\sigma$ = 0.6, 0.7, 0.8, 1.0, and 1.2. The dotted lines are denoted by $y_{b}=2h/\zeta_{0}$ with $\zeta_{0}$ = 1, 2, $\cdots$, 8.}
\end{figure}

Figures \ref{fig05}(a) and (b) show a plot of $y_{b}$ as a function of $h$ ($y_{b}$ vs $h$ diagram) at various width $\sigma$. For small $\sigma$ (typically $\sigma\le 0.11$), $y_{b}$ monotonically decreases with increasing $h$. At a larger $\sigma$ (typically $0.6 \le \sigma \le 1.2$), $y_{b}$ decreases with increase in $h$ for very low $h$, and shows a local minimum around $h = 0.05$. With further increasing $h$, $y_{b}$ increases with increasing $h$, showing a local maximum around $h = 0.10 - 0.15$, and decreases with further increasing $h$. The local maximum position shifts to high-$h$ side with increasing $\sigma$. The origin of the local maximum in the $y_{b}$ vs $h$ curves arises mainly due to the nonlinearity of the Langevin function in Eq.~(\ref{eqn35}), as will be discussed in Sec.~\ref{dis}. It should be noted that this diagram of $y_{b}$ vs $h$ may be a feature common to any SPM's, since it depends only on the scaled variables $h$ and $y$.

\begin{figure}
\includegraphics[width=7.0cm]{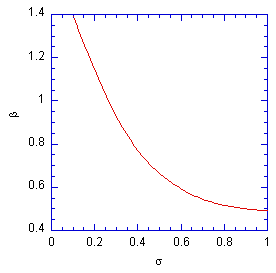}
\caption{\label{fig06}(Color online) Exponent $\beta$ vs $\sigma$. The exponent $\beta$ is derived from the least-squares fit of the $y_{b}$ vs $h$ curves for $0.1\lesssim h\lesssim 0.6$ to Eq.~(\ref{eqn37}). $\alpha = 2.0$.}
\end{figure}

As shown in Fig.~\ref{fig05}(a), the curvature of $y_{b}$ vs $h$ diagram drastically changes with varying $\sigma$. We assume that the $h$ dependence of $y_{b}$ is expressed by a power law form 
\begin{equation}
h=h_{0} (1-\frac{y_{b} }{y_{0} } )^{\beta} ,
\label{eqn37}
\end{equation}
for $0.1\le h\le 0.6$, where $\beta$ is an exponent and $h_{0}$ and $y_{0}$ are constants. The least-squares fit of the numerical result of $h$ vs $y_{b}$ for each $\sigma$ to Eq.~(\ref{eqn37}) yields the parameters $\beta$, $y_{0}$, and $h_{0}$. In Fig.~\ref{fig06}, we show the value of $\beta$ thus obtained, as a function of $\sigma$. We find that the exponent $\beta$ is strongly dependent on the width $\sigma$: $\beta$ increases with decrease in $\sigma$. The exponent $\beta$ is nearly equal to 0.5 at $\sigma$ = 1.2. It increases with decreasing $\sigma$ and is nearly equal to 1.5 at $\sigma$ = 0.01. Note that $\beta = 3/2$ is an exponent predicted for the de Almeida-Thouless (AT) line \cite{ref21} for the $H$-$T$ diagram in Ising SG systems. 

\begin{figure}
\includegraphics[width=7.0cm]{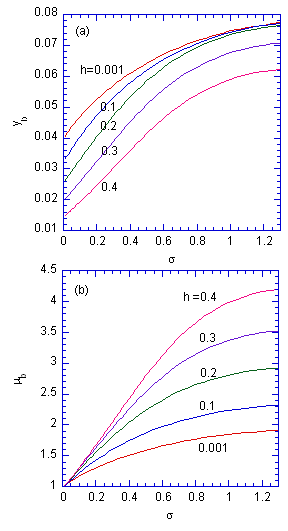}
\caption{\label{fig07}(Color online) (a) $y_{b}(\sigma, h)$ vs $\sigma$ at $h$ = 0.001, 0.1, 0.2, 0.3, and 0.4. $\alpha$ = 2.0. (b) Plot of $\mu_{b} = y_{b}(\sigma, h)/y_{b}(\sigma \rightarrow 0, h)$ vs $\sigma$ at various $h$. $y_{b}(\sigma, h)$ is the normalized blocking temperature at fixed $h$ and $\sigma$.}
\end{figure}

Figure \ref{fig07}(a) shows the plot of $y_{b}(\sigma , h)$ as a function of $\sigma$ for a fixed $h$ ($= 0.001 - 0.4$). The value of $y_{b}$ increases with increasing $\sigma$ at each $h$. Figure \ref{fig07}(b) shows the ratio $\mu_{b}$ defined as $\mu_{b} = y_{b}(\sigma , h)/y_{b}(\sigma \rightarrow 0, h)$ as a function of $\sigma$ at various $h$, where $y_{b}(\sigma \rightarrow 0, h)$ is obtained from the extrapolation of $y_{b}(\sigma , h)$ in the limit of $\sigma \rightarrow 0$. We find that the ratio $\mu_{b}$ is proportional to $\sigma$ at small $\sigma$ and tends to saturate at large $\sigma$. Our result is rather different from the prediction by Sappey et al. \cite{ref08} above $\sigma$ = 0.8. Their value of $\mu_{b}$ tends to diverge with increasing $\sigma$ above $\sigma$ = 0.8, while that of ours tends to saturate. The reason for such a difference in $\mu_{b}$ at large $\sigma$ is that the nonlinearity of the Langevin function with $y$ and $h$ is not taken into account in the paper by Sappey et al. \cite{ref08}. 

\begin{figure}
\includegraphics[width=7.0cm]{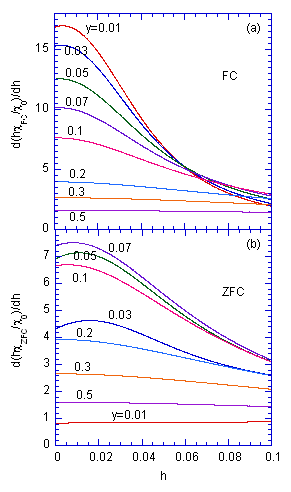}
\caption{\label{fig08}(Color online) $h$-dependence of (a) d$(h\chi_{FC}/\chi_{0})$/d$h$ and (b) d$(h\chi_{ZFC}/\chi_{0})$/d$h$ at various $y$ (= 0.01 - 0.5). $\sigma$ = 0.6. $\alpha$ = 2.0. $y_{b}$ is nearly constant ($\approx 0.065$) for $0.02\le h\le 0.12$ (see Fig.~\ref{fig05}).}
\end{figure}

\begin{figure}
\includegraphics[width=7.0cm]{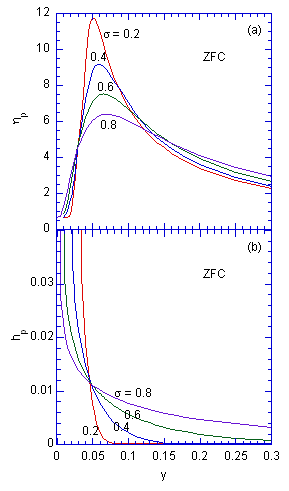}
\caption{\label{fig09}(Color online) (a) The peak height $\eta_{p}$ and (b) the peak field $h_{p}$ as a function of $y$ at various $\sigma$, where d$(h\chi_{ZFC}/\chi_{0})$/d$h$ vs $h$ curve exhibits a peak height $\eta_{p}$ at a peak field $h_{p}$ for each $y$. $\alpha=2.0$.}
\end{figure}

Figure \ref{fig08}(a) shows the $h$ dependence of the derivative d$(h\chi_{FC}/\chi_{0})$/d$h$ at various $y$, where $\sigma=0.6$, $\alpha=2.0$, and $(h\chi_{FC}/\chi_{0})$ is proportional to $M_{FC}$. The value of $y_{b}$ is nealy constant ($\approx 0.065$) for $0.02\le h\le 0.12$ (see Fig.~\ref{fig05}). For $y<y_{b}$, it shows a sharp peak around $h=0$. For $y>y_{b}$, d$(h\chi_{FC}/\chi_{0})$/d$h$ becomes flat around $h=0$. The pinched hysteresis form for $M_{FC}$ vs $H$ near $H=0$ observed for SPM's \cite{ref05,ref07,ref09} implies that the derivative of $M_{FC}$ with respect to $H$ exhibits a sharp peak. Thus these experimental results can be explained by the present model. Figure \ref{fig08}(b) shows the $h$ dependence of the derivative d$(h\chi_{ZFC}/\chi_{0})$/d$h$ at various $y$, where $\sigma=0.6$ and $(h\chi_{ZFC}/\chi_{0})$ is proportional to $M_{ZFC}$. It shows a broad peak at $h = h_{p}$ (= 0.0159) for $y=0.03$. This peak field $h_{p}$ shifts to the lower-$h$ side with increasing $y$, while the peak height $\eta_{p}$ increases with increasing $y$, showing a local maximum at $h_{p} = 0.0089$ for $y \approx 0.065$. We note that these values of $y$ and $h_{p}$ for the local maximum are located on the $y_{b}$ vs $h$ diagram $(h\simeq 0)$ for $\sigma$ = 0.6 (see Fig.~\ref{fig05}(b)). Figures \ref{fig09}(a) and (b) show the peak height $\eta_{p}$ and the peak field $h_{p}$ as a function of $y$ at various $\sigma$ for d$(h\chi_{ZFC}/\chi_{0})$/d$h$ vs $h$, respectively. The peak height $\eta_{p}$ shows a local maximum around $y = 0.05 - 0.06$. The peak field $\eta_{p}$ decreases with increasing $y$. These features are independent of $\sigma$.

In summary, we obtain the following important features from the above numerical calculations. (i) The normalized ZFC susceptibility $\chi_{ZFC}/\chi_{0}$ vs $y$ has a broad peak at $y = y_{b}$, forming the $y_{b}$ vs $h$ diagram. The value of $y_{b}$ exhibits a local minimum and a local maximum for large $\sigma$ ($\sigma >0.4$). (ii) The derivative d$(h\chi_{FC}/\chi_{0})$/d$h$ vs $h$ shows a very sharp peak around $h$ = 0 at low $y$. (iii) The derivative d$(h\chi_{ZFC}/\chi_{0})$/d$h$ vs $h$ shows a peak at the boundary of the $y_{b}$ vs $h$ diagram ($h\simeq 0$). (iv) The monotonic increase of $\chi_{FC}/\chi_{0}$ with decreasing $y$ is seen below $y_{b}$ for relatively large $\sigma$. With decreasing $\sigma$, $\chi_{FC}/\chi_{0}$ becomes flat below $y_{b}$ like SG's. 

\section{\label{dis}Discussion}
Both the non-monotonic $H$ dependence for the ZFC-peak temperature $T_{b}$ \cite{ref02,ref07,ref08,ref11,ref12} and the sharp peak of the derivative d$M$/d$H$ have been observed in many SPM's \cite{ref05,ref07,ref09}. Several theories have been presented for the explanation of the maximum of $T_{b}$ at low $H$ \cite{ref08,ref11,ref15} including the theories based on the N\'{e}el-Brown model \cite{ref13,ref14} and the resonant spin tunneling theory \cite{ref06,ref15}. The peak of $T_{b}$ at small $H$ in Mn$_{12}$O$_{12}$(CH$_{3}$COO)$_{16}$(H$_{2}$O)$_{4}$ (denoted as Mn$_{12}$) \cite{ref06} and ferritin \cite{ref09} may be explained in terms of thermally assisted, field-tuned resonant tunneling between particles. However, this model may not be valid for the explanation of similar behaviors in SPM's with $T_{b}$ which is too high for the quantum effect to appear.

In Sec.~\ref{calc} we show that these features can be well explained in terms of the scaling form of $\chi_{ZFC}/\chi_{0}$ and $\chi_{FC}/\chi_{0}$. The local minimum and local maximum of $y_{b}$ vs $h$ tends to disappear as $\sigma$ becomes smaller. The non-monotonic behavior of $y_{b}$ vs $h$ is mainly due to the non-linearity of the Langevin function. The curvature of $y_{b}$ vs $h$ is strongly dependent on $\sigma$. It is interesting to discuss where the nonlinearity of the Langevin function is significant in the $y_{b}$ vs $h$ diagram. Note that $\chi_{ZFC}/\chi_{0}$ is given by Eq.~(\ref{eqn35}) using the Langevin function which is a function of
\begin{equation}
\zeta =(M_{s}V)H/(k_{B}T)=2hx/y,
\label{eqn38}
\end{equation}
where $x=V/\langle V \rangle$, $y=k_{B}T/(K_{u}\langle V \rangle)$, $h=H/H_{k}$, and $H_{k}=2K_{u}/M_{s}$. $M_{s}V$ is the magnetic moment of the particle of volume $V$ and $M_{s}$ is its magnetization per unit volume. The nonlinearity is considered to appear when $\zeta >1$. For convenience, we assume that $x = 1$, which means $V = \langle V \rangle$. Then $\zeta$ is approximated as $\zeta = \zeta_{0} = 2h/y$. In Fig.~\ref{fig05}(a), we make a plot of the straight lines denoted by the relation $\zeta_{0}=2h/y$, where $\zeta_{0}=1- 8$. It is clear that the local maximum and local minimum for each $\sigma$ are located on the lines with $\zeta_{0} = 1$ and $\zeta_{0}=3-4$, respectively. This implies that the nonlinearity of the Langevin function is crucial to the occurrence of the local maximum and local minimum. In the numerical calculation, Sappey et al. \cite{ref08} have used the approximation of the Langevin function for $\zeta \ll 1$, which may lead to results inconsistent with our numerical results. We note that $y_{b}$ with $\sigma >0.4$ exhibits a local minimum around $h = 0$ as shown in Fig.~\ref{fig05}(a). This local minimum is located around the line given by $\zeta_{0}=1$ in the $y_{b}$ vs $h$ curve with each $\sigma$. Experimentally Luis et al. \cite{ref10} have reported that in natural horse-spleen ferritin the curve of $T_{b}$ vs $H$ exhibits a local minimum at $H$ = 0.5 kOe and a maximum at $H$ = 3 kOe. This result is qualitatively consistent with the results of our numerical calculation.

\section{Conclusion}
We show that remarkable features in the $T$ and $H$ dependence of the ZFC and FC susceptibility for SPM's (nanopraticles) can be well explained by the present model. These behaviors arises mainly from the nonlinearity of Langevin function, but not due to the quantum tunneling effect.\\
\\
\\
\textbf{Acknowledgments}\\

We are grateful to Prof. C. J. Zhong and Dr. L. Wang for useful discussion on the structures of nanoparticles in SPM's.

\end{document}